**Magnetic structure and magnetoelectric coupling in antiferromagnet $Co_5(TeO_3)_4Cl_2$**

B. Yu,[1] L. Huang,[1] J. S. Li,[2] L. Lin,[2,1*] V. Ovidiu Garlea,[3] Q. Zhang,[3] T. Zou,[4] J. C. Zhang,[5] J. Peng,[5] Y. S. Tang,[6] G. Z. Zhou,[1] J. H. Zhang,[1] S. H. Zheng,[7] M. F. Liu,[7] Z. B. Yan,[1] X. H. Zhou,[1] S. Dong,[5] J. G. Wan,[1,†] and J.-M. Liu[1]

[1]*Laboratory of Solid State Microstructures, Nanjing University, Nanjing 210093, China*

[2]*Department of Applied Physics, College of Science, Nanjing Forestry University, Nanjing 210037, China*

[3]*Neutron Science Division, Oak Ridge National Laboratory, Oak Ridge, Tennessee 37831, USA*

[4]*Collaborative Innovation Center of Light Manipulations and Applications, Shandong Normal University, Jinan 250358, China*

[5]*Key Laboratory of Quantum Materials and Devices of Ministry of Education, School of Physics, Southeast University, Nanjing 211189, China*

[6]*School of Science, Nanjing University of Posts and Telecommunications, Nanjing 210023, China*

[7]*Institute for Advanced Materials, Hubei Normal University, Huangshi 435002, China*

[*] llin@njfu.edu.cn

[†] wanjg@nju.edu.cn

**[Abstract]** The van der Waals (vdW) layered multiferroics, which host simultaneous ferroelectric and magnetic orders, have attracted attention not only for their potentials to be utilized in nanoelectric devices and spintronics, but also offer alternative opportunities for emergent physical phenomena. To date, the vdW layered multiferroic materials are still very rare. In this work, we have investigated the magnetic structure and magnetoelectric effects in $Co_5(TeO_3)_4Cl_2$, a promising new multiferroic compound with antiferromagnetic (AFM) Neel point $T_N \sim 18$ K. The neutron powder diffraction reveals the non-coplanar AFM state with preferred Néel vector along the *c*-axis, while a spin re-orientation occurring between 8 K and 15 K is identified, which results from the distinct temperature dependence of the non-equivalent Co sites moment in $Co_5(TeO_3)_4Cl_2$. What's more, it is found that $Co_5(TeO_3)_4Cl_2$ is one of the best vdW multiferroics studied so far in terms of the multiferroic performance. The measured linear ME coefficient exhibits the emergent oscillation dependence of the angle between magnetic field and electric field, and the maximal value is as big as 45 ps/m. It is suggested that $Co_5(TeO_3)_4Cl_2$ is an appreciated platform for exploring the emergent multiferroicity in vdW layered compounds.

## I. Introduction

Multiferroic materials, which simultaneously possess more than one ferroic order, have attracted attention due to their intriguing physical phenomena and potential applications [1-3]. In particular, the cross-coupling between magnetic and electric orders allows a control of magnetization ($M$) by electric field ($E$) or electric polarization ($P$) by magnetic field ($H$) [4-6]. However, most of the observed spin order induced ferroelectricity effects occur at low temperature ($T$) or the magnetoelectric (ME) coupling is more or less weak [7,8], leaving linear ME materials opportunity to be further explored, such as the time-honored $Cr_2O_3$ [9-11]. Based on the symmetry requirement, these linear ME compounds usually exhibit collinear spin order with non-polar magnetic point group and non-ferroelectricity. The magnetically induced ferroelectricity can only be triggered by magnetic field ($H$).

It is noted that linear ME effect does not necessarily require a highly frustrated structure and may be allowed even in non-polar crystals [10,12,13], offering many more opportunities for finding linear ME compounds including two-dimensional (2D) materials. Indeed, a set of emergent ferroic phenomena have been unveiled recent years in 2D van der Waals (vdW) family of materials such as ferroelectricity in $CuInP_2S_6$ [14] and $In_2Se_3$ [15,16], magnetism in $Cr_2Ge_2Te_6$ [17] and $CrI_3$ [18,19] etc. Although $CuCrP_2S_6$ [20-22] and $NiI_2$ were reported to exhibit comparatively ME effect [23-26], 2D vdW multiferroics are yet rare. In such sense, searching for 2D vdW multiferroic compounds seems to be a cutting-edge topic these years.

Recently, vdW magnets $A_5(TeO_3)_4X_2$ (A= Ni, Co, and $X$ = Cl, Br, I) in monoclinic structure with space group $C2/c$ have been receiving attention for their attractive magnetic structures and lattice symmetry variants. The crystal structure of this family consists of three unequal A-octahedra [$A(1)O_6$], [$A(2)O_5X$] and [$A(3)O_6$], forming a clamp-like corner shared [$A_5O_{17}X_2$] units [27,28]. Each layers consisting of two clamp-like [$A_5O_{17}X_2$] units with center of inversion symmetry are stacked along the $a$-axis through vdW interaction to form a layered structure. As revealed by neutron diffraction on $Ni_5(TeO_3)_4Br_2$ in Ref [29], the three inequivalent A-octahedra have different magnetic moments due to their different distortion degrees and crystal fields. Consequently, a non-collinear spin arrangement in the $ac$ plane below the Néel temperature $T_N$ = 29 K is established, resulting in a $C2'/c$ magnetic space group. Based on symmetry analysis, $C2'/c$ magnetic space group allows linear ME coupling, thus making $A_5(TeO_3)_4X_2$ family potentially a new class of multiferroics. Along this line, a success in observing the linear ME effect in $A_4B_2O_9$ (A = Mn, Fe, Co, Ni, B = Nb, Ta) compounds stimulates us to further explore those isostructural counterparts in $A_5(TeO_3)_4X_2$ family, noting that the spin-orbit coupling (SOC) plays a significant role in the origin of magnetically induced electric polarization [30-32]. Among the $A_5(TeO_3)_4X_2$ family, Co-based materials with $3d^7$ ($S$ = 3/2) magnetic cations possess stronger anisotropy and SOC effect [28]. However, to the best of our knowledge, the magnetic structures and ME coupling effects of $Co_5(TeO_3)_4X_2$ ($X$=Cl, Br, I) have not been well studied yet, deserving for further investigation.

Here, we experimentally demonstrate a large magnetic control of electric polarization in vdW magnet $Co_5(TeO_3)_4Cl_2$ ceramic sample and single crystal by a series of characterizations on structure, magnetic susceptibility, and ME response. Unlike the Ni-based compounds where the spins are confined within the *ac* plane, our powder neutron scattering data show the non-coplanar antiferromagnetic (AFM) order with Néel vector aligned along the *c*-axis. Furthermore, it is demonstrated that the samples exhibit larger magnetically induced electric polarization and stronger ME coupling in comparison with the currently known van der Waals layered magnetoelectric materials, suggesting that $Co_5(TeO_3)_4Cl_2$ would be a more promising new vdW multiferroic and a preferred platform for exploring 2D multiferroicity.

## II. Experimental details

Polycrystalline samples of $Co_5(TeO_3)_4Cl_2$ were prepared using conventional solid-state reaction method [28]. Stoichiometric mixture of high purity CoO, $CoCl_2$, and $TeO_2$ was thoroughly ground, and placed in evacuated silica ampoules, and sintered at 570 °C for 72 h in a muffle furnace. To ensure the sample purity, the aforementioned process was repeated for three times. Then, the pure-phase powder was thoroughly reground and pressed into a rod under a hydrostatic pressure. The rod-like sample was sintered at 570 °C for 72 h and cut into disks with ~ 0.4 mm in thickness for subsequent electrical measurements. The single crystals were grown by chemical vapor transport technique with polycrystalline powder and $TeCl_4$ as transport agent as described in Ref. [28]. The samples were characterized by the X-ray diffraction (XRD, D8 Advanced, Bruker) in the $\theta$-$2\theta$ mode with Cu $K_\alpha$ source ($\lambda$ = 1.5406 Å) at room temperature. To verify the stoichiometry of samples, the chemical composition was analyzed using electron dispersion spectroscopy (EDS) with a FEI Quanta 200 Scanning Electron Microscope (SEM). Additionally, the morphology of samples was examined using SEM to highlight the preferred orientation.

The magnetic susceptibility ($\chi$) as a function of $T$ from 2 K to 300 K was measured using the Quantum Design Superconducting Quantum Interference Device magnetometer (SQUID) in both the zero-field-cooled (ZFC) and field-cooled (FC) modes, with measuring field $\mu_0 H = 1$ kOe. The $H$-dependent magnetization $M$ was measured at selected $T$ using the vibrating sample magnetometer in the Physical Property Measurement System (PPMS, Quantum Design). The specific heat ($C_P$) was measured using the PPMS in the standard procedure.

Neutron powder diffraction was conducted at powder diffractometer (POWGEN) located at the Spallation Neutron Source (SNS), Oak Ridge National Laboratory. The POWGEN automatic sample changer (PAC) was selected as the sample environment to cover the $T$-range from 10 K to 300 K. The powder sample was loaded in a standard vanadium can and sealed with the helium exchange gas. The neutron bank with center wavelength 2.665 Å was used to collect the neutron data covering $d$ spacing from 1 Å to 21 Å. Magnetic structure was determined by the Rietveld analysis on the POWGEN data using GSAS-II software [33].

To measure ferroelectric properties, a sandwich-type capacitor was fabricated by depositing a layer of Au on both the top and bottom surfaces of a sample with a diameter of 5.0 mm and a thickness of ~ 0.4 mm. The pyroelectric current was measured from 2 K to 30 K with a $T$-ramp rate of 4 K/min using a Keithley 6514 electrometer. To ensure a single-domain state, the ME poling procedure was applied. In details, the sample was poled under an electric field of 3.33 kV/cm and selected magnetic field over $T$-range from 50 K to 2 K. Then the poling electric field was removed before measuring the pyroelectric current. To investigate the ME effect, the $H$-dependent ME current ($I_{\mathrm{ME}}$) was measured under selected $T$ upon $\mu_0 H$ ramping from +9 to -9 T at a rate of 100 Oe/s using the same ME poling procedure. Finally, the $T$ and $H$ dependences of electric polarization $P$ was obtained by integrating of the pyroelectric current or magnetoelectric current with time. For the ME measurements of single crystal, the sample was applied under the same ME poling procedure in the $E$ // $a^*$ axis and $H$ parallel to $a^*$, $b$, and $c$ axis, whereas $a^* \perp bc$ plane.

## III. Results and discussion

### *A. Crystal structure*

As mentioned, $Co_5(TeO_3)_4Cl_2$ belongs to monoclinic structure with space group $C2/c$ and is isomorphic to $Ni_5(TeO_3)_4Cl_2$. Figure 1(a) illustrates how $Co_5(TeO_3)_4Cl_2$ units are stacked along the *a*-axis through inter-laminar vdW interactions. Three unequal Co-octahedra [Co(1)$O_6$], [Co(2)$O_5$Cl] and [Co(3)$O_6$], whereas symbols $Co_1$, $Co_2$, and $Co_3$ denote the different Co occupations forming a clamp-like Cobalt pentamer cluster of [$Co_5O_{17}Cl_2$] units. In Fig. 1(b) we present the slow-scan XRD pattern of the ceramic sample. The robust orientation preference is observed, evidenced by the strong intensity of the pronounced sharp (*h*00) peaks, e.g., (200), (400), and (800), while the peaks in other orientation appear considerably much weaker. Further Rietveld refinement is performed on the fully ground crushed ceramics [28]. It shows that all the reflection peaks can be well fitted by standard database for *C*2/*c* monoclinic lattice and no impurity phase is detected. The refined structural parameters are $a$ = 19.729(2) Å, $b$ = 5.264(8) Å, $c$ = 16.437(1) Å, $\alpha = \gamma = 90$ °, and $\beta$ = 125.25(9) °, consistent with the reference [28]. To check more details of the grains and crystal orientation, we performed SEM image of the ceramic sample as shown in Figs. 1(c), which exhibits a prominent flake shape suggesting a strong orientation preference. Furthermore, a few inclined and vertical sheets can be observed, contributing to the strong intensity of XRD pattern. The analysis of sample chemical composition using EDS reveals an atomic ratio of Co: Te: Cl as 5.19: 4.03: 2, which is consistent with the expected stoichiometric ratio of $Co_5(TeO_3)_4Cl_2$, as shown in Fig. 1(d).

*B. Magnetic properties and specific heat*

Figure 2(a) depicts the $T$ dependence of magnetic susceptibility ($\chi$) in the ZFC and FC conditions. The overlapping ZFC and FC curves both show a sharp peak at $T = T_N \sim 18$ K, the AFM Neel point. The Curie-Weiss fitting of the data from 100 K to 300 K gives rise to Curie-Weiss temperature $\theta_{CW} \sim -68$ K, indicating the AFM nature of the $Co^{2+}$ spin exchange. The derived effective paramagnetic moment $\mu_{eff} \sim 5.2$ $\mu_B/Co^{2+}$ falls in the expected spin-only moment 3.87 $\mu_B$/Co and total effective moment 6.64 $\mu_B$/Co, revealing the contribution from the orbital angular momentum via the spin-orbit coupling. In addition, the frustration factor $f = |\theta_{CW}/T_N|$ is ~ 3.8, implying a weak frustration. The measured isothermal $M$-$H$ dependences at different $T$ show no evident field-induced metamagnetic transition up to $\mu_0 H \sim 9$ T as shown in Fig. 2(b).

The AFM ordering at $T_N$ can be further confirmed by the heat capacity data, as shown in Fig. 2(c), exhibiting the typical $\lambda$-like anomaly at $T_N$. While the standard Debye model is used to evaluate the phonon contribution $C_{Latt}$, a subtraction of this term allows the magnetic contribution $C_M$ to be obtained, as shown by the blue line in Fig. 2(c). The magnetic entropy $\Delta S_M$ estimated by integrating $C_M/T$ over the whole $T$-range, presented by the blue line in Fig. 2(d), shows a value of ~ 37 J mol$^{-1}$ K$^{-1}$ of magnetic entropy, about 60% of the total value 5Rln(2$J$ + 1) = 57.6 J mol$^{-1}$ K$^{-1}$, indicating the $Co^{2+}$ angular momentum $J = 3/2$.

*C. Neutron powder diffraction*

To unveil the spin configuration with the AFM state, we discuss the neutron powder diffraction (NPD) data shown in Fig. 3, collected at $T$ = 30 K, 15 K, and 8 K, respectively. The Rietveld refinement confirms the space group $C2/c$. At $T$ = 30 K, the refined lattice constants are $a$ = 19.557(7) Å, $b$ = 5.259(3) Å, and $c$ = 16.366(4) Å, consistent with the XRD data and details are listed in Table I. It is noted that additional Bragg peaks, e.g., Q ~ 0.47 and 0.64 Å$^{-1}$ appear below $T_N$, due to the magnetic scattering. By subtracting the paramagnetic nuclear contribution ($T$ = 30 K) from the data obtained at $T$ = 15 and 8 K, the pure magnetic contribution was isolated, as shown in the insets of Figs. 3(b) and 3(c). The clear magnetic reflections denoted by (001), (20-1), and (20-3) are well explained by the wave propagation vector $\boldsymbol{k}$ = (0, 0, 0).

To gain deeper insights into the magnetic structure, it is instructive to map out the magnetic Co sublattices. Fig. 4 (a) shows the three unequal Co-octahedra [$Co(1)O_6$], [$Co(2)O_5Cl$] and [$Co(3)O_6$], whereas symbols $Co_1$, $Co_2$, and $Co_3$ denote the different Co sites forming a clamp-like Cobalt pentamer cluster of [$Co_5O_{17}Cl_2$] units at $T > T_N$. These clusters consist of five Co-octahedra, denoted as $Co_1$, $Co_1$', $Co_2$, $Co_2$', and $Co_3$, as illustrated in Figs. 4(b) to 4(e). The arrangement of the three non-equivalent Co sites is reminiscent of the magnetically frustrated bow-tie lattice consisting of vertex-sharing triangles.

Based on the symmetry analysis using the tools at Bilbao Crystallographic Server [34], the irreducible representation *mGM2* - corresponding to the magnetic space group $C2'/c$ yields the closest agreement with the experimental data. The magnetic point group is $2'/m$ below $T_N$, which forbids spontaneous polarization but allows the linear ME effect. The magnetic moments of $Co_1$ and $Co_2$ positions are allowed to point in any direction of the lattice, while the $Co_3$ moment is constrained to lie in the *ac*-plane. The refined values for $Co^{2+}$ moments at each site and at $T$ = 15 and 8 K are given in Table II. It is found that the $Co_3$ ion situated at the center bow-tie has a much smaller ordered moment, which indicates a strong site-dependent magnetic frustration, similar to the observations for similar bow-tie system $Ca_2Mn_3O_8$ [35] and $Mn_5(VO_4)_2(OH)_4$ [36]. The $Co^{2+}$ moments form a non-coplanar structure, in which the nearest $Co_1$ and $Co_2$ moments tend to align in the same direction, but opposite to the $Co_3$ moment.

In addition, a comparison of the magnetic arrangements refined at 15 and 8 K confirms that the temperature dependence of the $Co_3$ moment is different from those of $Co_1$ and $Co_2$. The $Co_3$ moment value increases more rapidly (77%) compared to the other two moment values (61% for $Co_1$ and 57% for $Co_2$). It is worth noting that the magnetic moment at different sites varies significantly along different directions. The $Co_3$ moment value ($M_z$) along the *z* direction has increased by 77%, while the $Co_1$ and $Co_2$ moment values ($M_z$) have increased by 19% and 57%, respectively. The variation is just in accordance with a previous report in $Ni_5(TeO_3)_4Br_2$ by Pregelj *et al.*, whereas the distinct temperature dependence of the non-equivalent Ni sites has been revealed [29]. In contrast to $Ni_5(TeO_3)_4Br_2$, where the moments are exclusively confined to the *ac* plane, the magnetic moments of $Co_5(TeO_3)_4Cl_2$ are non-coplanar and include components along the *b*-direction. In addition, the $Co_2$ moment value ($M_y$) along the *y* direction decreases by 34%, while the sign of $Co_1$ moment $M_y$ has changed from *y* direction to -*y* direction, indicative of the spin reorientation between 8 K and 15 K.

*D. Anisotropic ME effects*

Herein special attention is given to checking the potential ME coupling effect based on the magnetic structure. As the bulk sample shows strong preference orientation along the *a*-axis, for comparison, the *H*-induced electric polarization *P* as a function of *T* was measured for *H* parallel ($H_{//}$) and perpendicular ($H_{\perp}$) to the poling electric field *E*. Figure 5(a) and 5(b) depict the pyroelectric current (*I*) and *P*, given of $Co_5(TeO_3)_4Cl_2$ in a magnetic field of $\mu_0H_{//} = 0$ and 9 T with different *E* settings. It is seen that $P = 0$ at $\mu_0H_{//} = 0$, reasonable due to the centrosymmetric space group. However, the pyroelectric current *I* rises rapidly as *T* increases to $T_N$, and forms a broad peak at non-zero *H*. The measured *P* is also proportional to *E*, indicating incomplete polarized ME domains. Besides, the symmetrical *I*(*E*) shows that the polarization can be switched by reversing *E*. More examination of the ME effect is presented in Figs. 5(c) and 5(d). While a clear linear ME effect is seen below $\mu_0H = 7$ T, a deviation from the linear *P*(*H*) dependence, as seen in the inset of Fig. 5(d). Overall, $P$ can be as large as ~ 160 μC/m$^2$ at $\mu_0H_{//}$ = 8 T and $T$ = 2 K, an appreciated ME effect considering the polycrystalline nature of the samples. Nevertheless, as the bulk sample shows strong preference orientation, one is allowed to check the electric polarization in the perpendicular magnetic field. Figures. 5(e) and 5(f) show the $T$ dependence of $I$ and $P$ in various $H_\perp$ configurations. It is seen that similar ME response of $Co_5(TeO_3)_4Cl_2$ against $H$ is obtained, as shown in the inset of Fig. 5(f). However, it should be noted that the intensity of polarization has doubled, indicating the presence of an anisotropic ME effect in our polycrystalline sample.

In order to further investigate its linear ME coupling effect, the $H$ dependence of ME current ($I_{ME}$) and field-induced polarization ($\Delta P$) measured in the $H$ // $E$ configuration at different temperatures upon $\mu_0H$ ramping from +9 T ~ -9 T ~ +9 T at a rate of 100 Oe/s are displayed in Figs. 6(a) and 6(b), respectively, whereas the green arrow indicates the scanning magnetic field direction. As the magnetic field scans from 9 T to 0 T, there is a sharp increase of $I_{ME}$ between 7.5 T and 9 T, accompanied by slowly released current with the decrease of magnetic field. The current shape in the positive and negative magnetic field regions shows good symmetry. Fig. 6(b) present the electric polarization $\Delta P$ in the sweeping magnetic field, and the relationship between polarization and the magnetic field showcases a clear linear trend below 7 T with the coefficient $\alpha_{//}$ = 18 ps/m. It is noted that the ME response shows obvious deviation from linearity at high field, and the critical field is roughly consistent with that observed in the iso-field experiments as shown in the inset of Fig. 5(f). Moreover, the ME response in the $H \perp E$ configuration exhibits a similar linear trend, as shown Fig. 6(d). However, the linear ME coupling coefficient $\alpha_\perp$ reaches an impressive magnitude of 45 ps/m, larger than $\alpha_{//}$ and the currently known van der Waals layered ME materials, such as $MX_2$ (M=transition metal, X= halogen) [37-41], and $CuCrP_2S_6$ [20]. More details of the ME performance of some vdW ME materials are summarized in the TABLE III.

To further illustrate the anisotropic ME effects, we performed the angular ($\theta$) dependence of magnetically induced polarization $\Delta P$ in rotating $\mu_0H$ = 7 T at $T$ = 2 K after the ME poling procedure. Here, the angle $\theta$ denotes the direction of magnetic field with respect to the direction of the normal vector on a sample surface, as shown in the Fig. 7. Thus $\theta$ = 0 and $\theta$ = 90° represent $H$ // $E$ and $H \perp E$ configurations. The ME coupling coefficient in the parallel and vertical configurations is denoted as $\alpha_{//}$ and $\alpha_\perp$, respectively. Due to the polycrystalline feature, the magnetically induced polarization $\Delta P$ is proportional to $\Delta P = \alpha_{//}H_{//} + \alpha_\perp H_\perp$. Interestingly, we clearly see a periodic variation of $\Delta P$ upon rotating $H$, a direct proof of dominating linear ME effect in this system. In this configuration, $H_{\perp} = H\sin\theta$ and $H_{//} = H\cos\theta$, and $\Delta P$ can be expressed as

$$\Delta P = H\left(\alpha_{//}\cos\theta + \alpha_{\perp}\sin\theta\right) = H\sqrt{\alpha_{\perp}^{2} + \alpha_{//}^{2}}\,\sin\left(\theta + \theta_{0}\right), \tag{1}$$

where $\theta_0$ = arctan ($\alpha_{//}/\alpha_{\perp}$). In Fig. 7, we plot the electric polarization as a function of $\theta$, whereas a periodic variation of $\Delta P$ can be observed. Here, the arrow indicates the direction of the scanning field (red arrow: 0°~360°, blue arrow: 360°~0°). It is noted that the difference in the polarization value at $\theta$ = 0° and at $\theta$ = 180° may be due to the different preferred orientation distributions of the ceramic sample. We used sinusoidal function A*sin[($\theta$-66)π/180] as the green dashed line to fit $\Delta P$, and $\theta_0 \approx$ 66° can be deduced, giving rise to $\alpha_{//}/\alpha_{\perp} \approx 2.2$. Intriguingly, this ratio is almost consistent with our experimentally observed $\alpha_{//}/\alpha_{\perp}$ = 45/18 $\approx$ 2.4 from the data in Fig. 6.

*E. Magnetic property and ME effect in* $Co_5(TeO_3)_4Cl_2$ *single crystal*

In order to clarify whether the measured ME anisotropy is solely due to the crystallographic orientation of the 'grains' or its intrinsic property, further in-depth discussion regarding the ME effect and spin reorientation demonstrated in polycrystalline sample is highly needed. We were fortunate to obtain a small amount of $Co_5(TeO_3)_4Cl_2$ single crystals, and try to gather more information on the characteristics of spin reorientation through magnetization measurements. As shown in Fig. 8, the naturally grown crystals exhibit transparent and purple flaky morphology with the size of 2×3 mm$^2$. We performed the room temperature slow scan XRD onto the naturally developed plane, and sharp peaks can be well indexed by the ($h$00) reflections. A narrow full width at half maximum (FWHM = 0.04°) of the typical (200) orientation is found, indicative of strict orientation growth and high crystallinity. These high quality single crystals offer the potential to conduct further detailed measurements. Fig. 9(a) depicts the $T$ dependence of magnetic susceptibility ($\chi$) in the ZFC condition along the $a^*$, $b$ and $c$ axes, where $a^* \perp bc$-plane. Upon further decreasing $T$, a broad peak appears at $T_N \sim$ 18 K which is assigned as antiferromagnetic Néel point, followed by a clear anomaly at $T_S \sim$ 13 K both along the $a^*$-axis and $b$-axis, indicating that re-arrangements of magnetic moment may occur below $T_S$. As shown in Fig. 9(b), the Curie-Weiss fitting of the data from 100 K to 300 K gives rise to the Curie-Weiss temperature $\theta_{a^*} \sim -61$

K, $\theta_b \sim -43$ K, and $\theta_c \sim -73$K (details are listed in Table IV), respectively, indicating the AFM nature of the $Co^{2+}$ spin exchange and magnetic anisotropy. The derived effective paramagnetic moment along $a$*-axis, $b$-axis, $c$-axis is $\mu_{eff} \sim 4.7\ \mu_B/Co^{2+}$, $5.0\ \mu_B/Co^{2+}$, $5.4\ \mu_B/Co^{2+}$, respectively, which falls in the expected spin-only moment 3.87 $\mu_B$/Co and total effective moment 6.64 $\mu_B$/Co, revealing the contribution from the orbital angular momentum via the spin-orbit coupling.

As shown in Fig. 9(c), the temperature dependence of pyroelectric current ($I$) along $a$* axis reveals a notable anomaly at $T_N$ when an applied magnetic field is aligned parallel to the $b$-axis. The pyroelectric current and electric polarization under different $H$ parallel to $b$ axis indicates magnetoelectric coupling, as shown in Figs. 9(c) and 9(d). The electric polarization $P$ can be as large as ~ 170 μC/m$^2$ at $H_b = 9$ T and $T = 2$ K, and the corresponding ME coefficient $\alpha_{xy} = 18$ ps/m. Here, we employ the IEEE standard setting, where we define the physical property axis ($x$, $y$, $z$), namely $x \perp yz$-plane, $y$//[010], and $z$//[001]. However, there are almost zero values in other magnetic directions, $\alpha_{xx} = \alpha_{xz} = 0$, which is consistent with the matrix form of the ME tensor for $C2'/c$ magnetic space group. Imposed by the magnetic point group 2′/m, in principle there are four nonzero terms of ME coupling tensor: $\alpha_{xy}$, $\alpha_{yx}$, $\alpha_{zy}$, and $\alpha_{yz}$. It is worth noting that due to the thin flak-like nature of the crystal along $a$* axis, it is not feasible to validate the magnetoelectric coupling coefficient in alternative directions. However, at the very least, the single crystal data can demonstrate non-diagonal magnetoelectric coupling. Generally, the polarization of a polycrystalline sample is smaller compared to that of a single crystal. However, the polarization in this case is larger than that of a single crystal, indicating that polarization along other directions, such as the $y$ and $z$ axes polarization, might have larger values.

*F. Discussion*

$Co_5(TeO_3)_4Cl_2$ has a layered crystal structure that is stacked along the $a$-axis by van der Waals (vdW) force. According to the NPD analysis, one unit cell contains four clamp-like cobalt pentamer cluster of [$Co_5O_{17}Cl_2$], which consists of three inequivalent $Co^{2+}$ sites, named as $Co_1$, $Co_{1'}$, $Co_2$, $Co_{2'}$, $Co_{3'}$ and $Co_3$ ($Co_1$ and $Co_{1'}$, $Co_2$ and $Co_{2'}$ are related by inversion symmetry). The combination of magnetic anomaly at $T_S \sim 13$ K and distinct temperature dependence of the non-equivalent Co sites demonstrates that re-arrangements of magnetic moment may occur below $T_S$. However, the neutron scattering shows that magnetic space group $C2'/c$ remains unchanged up to 8 K. As a result, there are no anomalies in the temperature-dependent pyroelectric current and polarization curves before and after the spin reorientation.

According to the symmetry analysis, the magnetic point group $2'/m$ forbids spontaneous polarization but allows linear ME effect. Here $2'$ is twofold rotation along the $y$ direction, and $m$ is a mirror perpendicular to the $y$-axis. Meanwhile, due to the hidden symmetry $-1'$, $Co_5(TeO_3)_4Cl_2$ does not exhibit macroscopic polarization. However, under an applied magnetic field, the breaking of 2' or $m$-symmetry allows for the occurrence of electric polarization. Our ME data for the single crystals provide clear evidence of the anisotropic ME coupling, which is consistent with the matrix form of the ME tensor. In linear magnetoelectric materials, the induced electric polarization (magnetization) is linearly proportional to the applied magnetic field (electric field) as expressed by $P_i = \alpha_{ij}H_j$ or $M_i = \alpha_{ji}E_j$, where $\alpha_{ij}$ is the first-order magnetoelectric coefficient. Typical linear magnetoelectric coupling materials, such as $Cr_2O_3$ [42] and $Co_4Nb_2O_9$ [30], exhibit the fascinating properties of magnetically controlled polarization and electrically controlled magnetism. Based on the symmetry analysis using the tools at Bilbao Crystallographic Server [34], the magnetic point group 2′/m also allows for four nonzero magnetoelectric tensor $\alpha_{xy}$, $\alpha_{yx}$, $\alpha_{zy}$, and $\alpha_{yz}$, which relates electric field with magnetization. In principle, there is significant potential for magnetism to be controlled by electrical means in $Co_5(TeO_3)_4Cl_2$. However, it is unfortunately not accessible at this moment, deserving for further investigation. The major difficulty lies in that the magnetization variation is expected to be very small and hard to detect while the sample in the SQUID chamber must be connected with a stage which can be electrically controlled: it is challenging for a bulk single crystal.

Now, we turn to briefly discuss the magnetically induced polarization from a microscopic perspective. There are three well-known mechanisms to explain the microscopic origin of spin-driven ferroelectricity. In the exchange striction mechanism, the electric polarization arises due to the symmetric exchange interaction. A prominent example of this is $Ca_3CoMnO_6$, which exhibits up-up-down-down collinear spin order [43]. On the contrary, the inverse Dzyaloshinskii-Moriya (DM) interaction or the spin current model is derived from the antisymmetric exchange interaction in noncollinear magnetic orders, such as cycloidal or transverse conical spin structures [42]. The metal-ligand hybridization $p$-$d$ mechanism gives the local electric polarization through the spin-orbit interaction in some low-symmetry system [44]. As shown in Fig. 4, The $Co^{2+}$ magnetic moments form a non-coplanar magnetic structure, where the moments are almost entirely confined to the *xz*-plane forming a cycloid with the scalar product of neighboring magnetic moments as $h = S_i \times S_j$ along *y* direction. Then a local electric polarization $P_z$ (or $P_x$) = $h \times q$ along *z* (or *x*) direction can be established, in which *q* is the magnetic modulation vector along *x* (or *z*) axis. On the other hand, it is worth reminding that in a clamp-like cobalt pentamer cluster of [$Co_5O_{17}Cl_2$], [Co(1)$O_6$], [Co(2)$O_5$Cl] and [Co(3)$O_6$] octahedron exhibit low symmetry and significant distortion, which may also give rise to a local polarization due to the *p-d* hybridization mechanism. Hence, further measurements in the single crystal and theoretical calculation would be highly required to elucidate the contribution of ME coupling mechanisms in $Co_5(TeO_3)_4Cl_2$.

## IV. Conclusion

To summarize, we have investigated the magnetic structure and anisotropic magnetoelectric coupling effects in the polycrystalline $Co_5(TeO_3)_4Cl_2$ by combining magnetic susceptibility, heat capacity, neutron-powder diffraction, and electric polarization measurements. Unlike the Ni-based compounds, the magnetic order in $Co_5(TeO_3)_4Cl_2$ is found to be to non-coplanar with the moments preferentially pointing along the *c*-axis. Furthermore, $Co_5(TeO_3)_4Cl_2$ is characterized by a significantly larger ferroelectric polarization and a more pronounced linear magnetoelectric coupling effect, compared to the van der Waals layered materials that have been discovered so far. The magnetically induced electric polarization is confirmed by the $C2'/c$ magnetic space group determined by neutron scattering. In addition, we found that the electric polarization also exhibited periodical retention under a rotation of the magnetic field, and the fitted ratio of linear ME coefficient $\alpha_{//}/\alpha_{\perp}$ is quite consistent with our experimental results. It is believed that the intriguing ME response to rotating magnetic field can be attributed to the continuous changes of antiferromagnetic moments. The establishment of $Co_5(TeO_3)_4Cl_2$ as a new member of the two-dimensional layered magnetoelectric coupling family provides insights into designing the significant ME coupling properties in other van der Waals layered magnet.

**Acknowledgement**

The authors would like to acknowledge the financial support from the National Natural Science Foundation of China (Grants No. 92163210, No. 12274231, No.12074111, No. 52272108, No. 12304124, and No. 12304119). A portion of this research used resources at the Spallation Neutron Source, a DOE Office of Science User Facility operated by the Oak Ridge National Laboratory.

**TABLE I**. Structural parameters for $Co_5(TeO_3)_4Cl_2$ refined from NPD data at 30 K. Space group: *C*2/*c*, *a* = 19.557(7) Å, *b* = 5.259(3) Å, *c* = 16.366(4) Å, $\alpha = \gamma = 90°$, and $\beta = 125.05(5)°$.

| Atom | Site | *x* | *y* | *z* | *U* |
|---|---|---|---|---|---|
| Co1 | 8*f* | 0.0081(7) | 0.2196(4) | 0.6169(4) | 0.00051 |
| Co2 | 8*f* | 0.0925(3) | 0.2167(5) | 0.4754(0) | 0.00051 |
| Co3 | 4*e* | 0 | 0.2389(2) | 1/4 | 0.00051 |
| Te4 | 8*f* | 0.1287(5) | 0.3130(4) | 0.1407(4) | 0.00251 |
| Te5 | 8*f* | 0.1491(8) | 0.2936(1) | 0.8782(9) | 0.00251 |
| Cl6 | 8*f* | 0.2419(2) | 0.3218(9) | 0.5977(7) | 0.00270 |
| O7 | 8*f* | 0.0415(4) | 0.1626(1) | 0.1429(1) | 0.00250 |
| O8 | 8*f* | 0.0663(0) | 0.4909(1) | 0.3692(3) | 0.00250 |
| O9 | 8*f* | 0.0721(2) | 0.3893(4) | 0.5690(6) | 0.00250 |
| O10 | 8*f* | 0.0787(6) | 0.0179(5) | 0.8557(2) | 0.00250 |
| O11 | 8*f* | 0.0986(1) | 0.1434(0) | 0.0228(0) | 0.00250 |
| O12 | 8*f* | 0.1153(4) | 0.3431(1) | 0.7491(7) | 0.00250 |
| $R_{WP}$= 5.093, *GOF* = 4.16 | | | | | |

**TABLE II**. Refined structural parameters and magnetic moments obtained from the NPD data taken at $T$=15 and 8 K.

| Temperature | Atom | Site | $m_x$ ($\mu_B$) | $m_y$ ($\mu_B$) | $m_z$ ($\mu_B$) | $m_{tot}$ ($\mu_B$) |
|---|---|---|---|---|---|---|
| 15 K | Co1 | 8*f* | 1.7921 | 0.0912 | 0.2530 | 1.8122 |
| | Co1' | 8*f* | 1.7921 | -0.0912 | 0.2530 | 1.8122 |
| | Co2 | 8*f* | -1.8199 | 0.8392 | -0.9069 | 2.1997 |
| | Co2' | 8*f* | -1.82199 | -0.8392 | -0.9069 | 2.1997 |
| | Co3 | 4*e* | 0.73296 | 0 | 0.0613 | 0.7355 |
| 8 K | Co1 | 8*f* | 2.7920 | -0.8026 | 0.3022 | 2.9207 |
| | Co1' | 8*f* | 2.7920 | 0.8026 | 0.3022 | 2.9207 |
| | Co2 | 8*f* | -3.1286 | 0.4152 | -1.4235 | 3.4622 |
| | Co2' | 8*f* | -3.1286 | -0.4152 | -1.4235 | 3.4622 |
| | Co3 | 4*e* | 1.2977 | 0 | 0.1082 | 1.3022 |

$R_{WP}$ = 6.136, *GOF* = 5.79 at 8 K, $R_{WP}$ = 5.242, *GOF* = 4.92 at 15 K.

**TABLE III**. The ME performance of experimentally observed van der Waals ME materials.

| Material | Spontaneous $P$ ( μC/m$^2$) | Max $P$ ( μC/m$^2$) | ME coefficients $\alpha$ (ps/m) | $T_{FE}$ (K) | Ref |
|---|---|---|---|---|---|
| $MnI_2$ | ~ 80 | ~140 ($\mu_0 H$ = 2 T) | ~ 30 | 3.9 | [39] |
| $CoI_2$ | ~ 8 | ~10 ($\mu_0 H$ = 2 T) | <1 | 10 | [40] |
| $NiI_2$ | ~ 50 | ~120 ($\mu_0 H$ = 14 T) | ~ 20 | 58 | [40] |
| $NiBr_2$ | ~ 20 | ~25 ($\mu_0 H$ = 2 T) | <1 | 23 | [41] |
| $CuCl_2$ | ~ 25 | ~25 ($\mu_0 H$ = 4 T) | <1 | 24 | [42] |
| $CuBr_2$ | ~ 8 | ~22 ($\mu_0 H$ = 9 T) | ~ 2 | 73 | [43] |
| $CuCrP_2S_6$ | 0 | <1 | <1 | 32 | [19] |
| $FeTe_2O_5Cl$ | ~ 3 | - | - | 11 | [26] |
| $Co_5(TeO_3)_4Cl_2$ | 0 | ~320 ($\mu_0 H$ = 9 T) | ~ 45 | 18 | This work |

**TABLE IV.** The Curie-Weiss temperature ($\theta_{cw}$), frustration factor ($f$) and effective moment ($\mu_{eff}$) for $Co_5(TeO_3)_4Cl_2$ in the geometry aligned parallel to *a**-axis, *b*-axis, *c*-axis and polycrystal, respectively.

| $Co_5(TeO_3)_4Cl_2$ | $T_N$ (K) | $\theta_{cw}$ (K) | $f$ | $\mu_{eff}$ ($\mu_B$) |
|---|---|---|---|---|
| *a** | 18 | -61 | 3.4 | 4.7 |
| *b* | 18 | -43 | 2.3 | 5.0 |
| *c* | 18 | -73 | 4.1 | 5.4 |
| polycrystal | 18 | -68 | 3.8 | 5.2 |

## Figures and captions

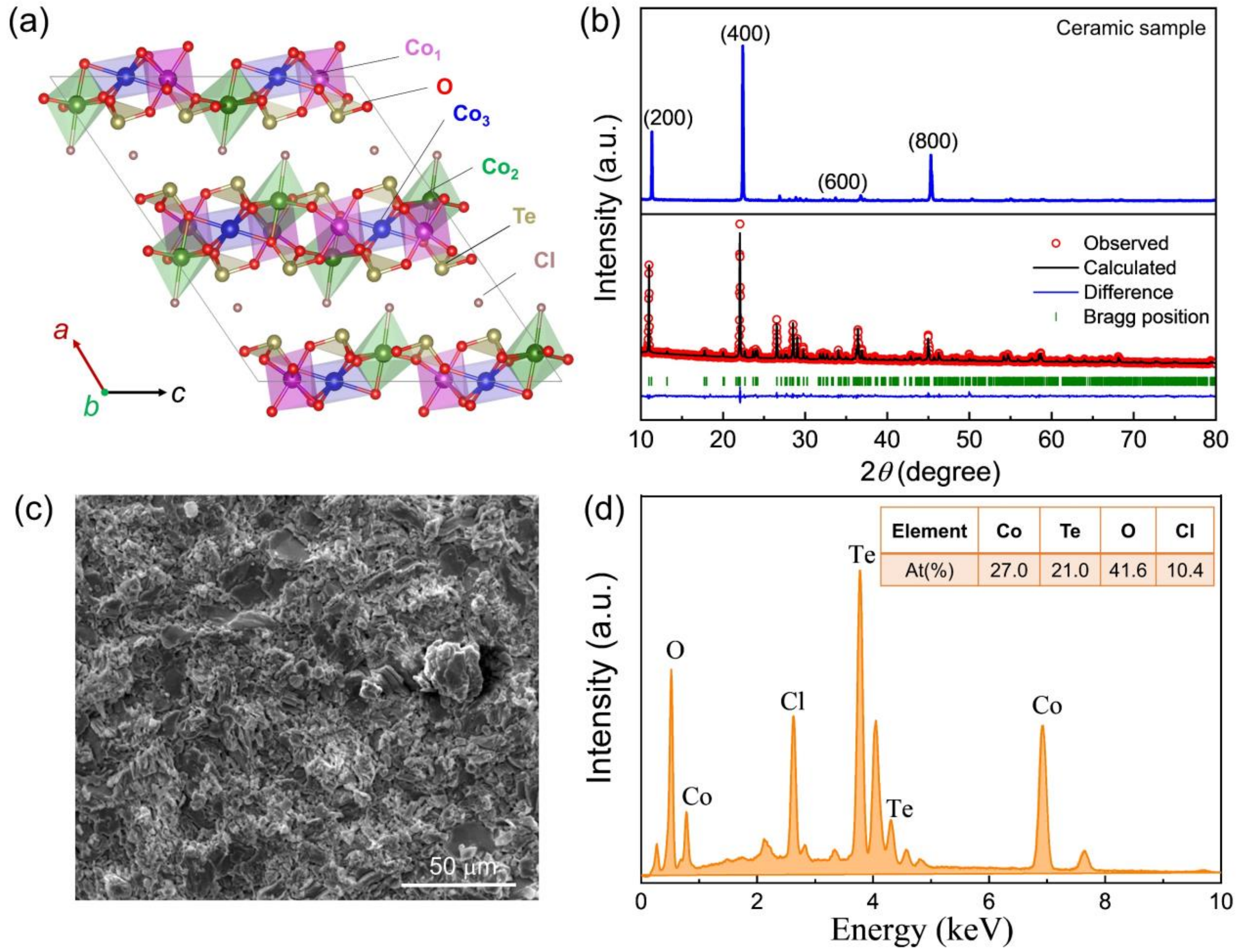

Fig. 1. (a) The crystal structure of $Co_5(TeO_3)_4Cl_2$ viewed along the [010] direction. Three cobalt--oxygen (red spheres) cages are illustrated in the magenta, green, and blue, respectively. (b) The XRD of ceramic sample and the refined XRD pattern of the crushed ceramics collected at room temperature. (c) Micromorphology of ceramic samples under SEM and (d) EDS spectra.

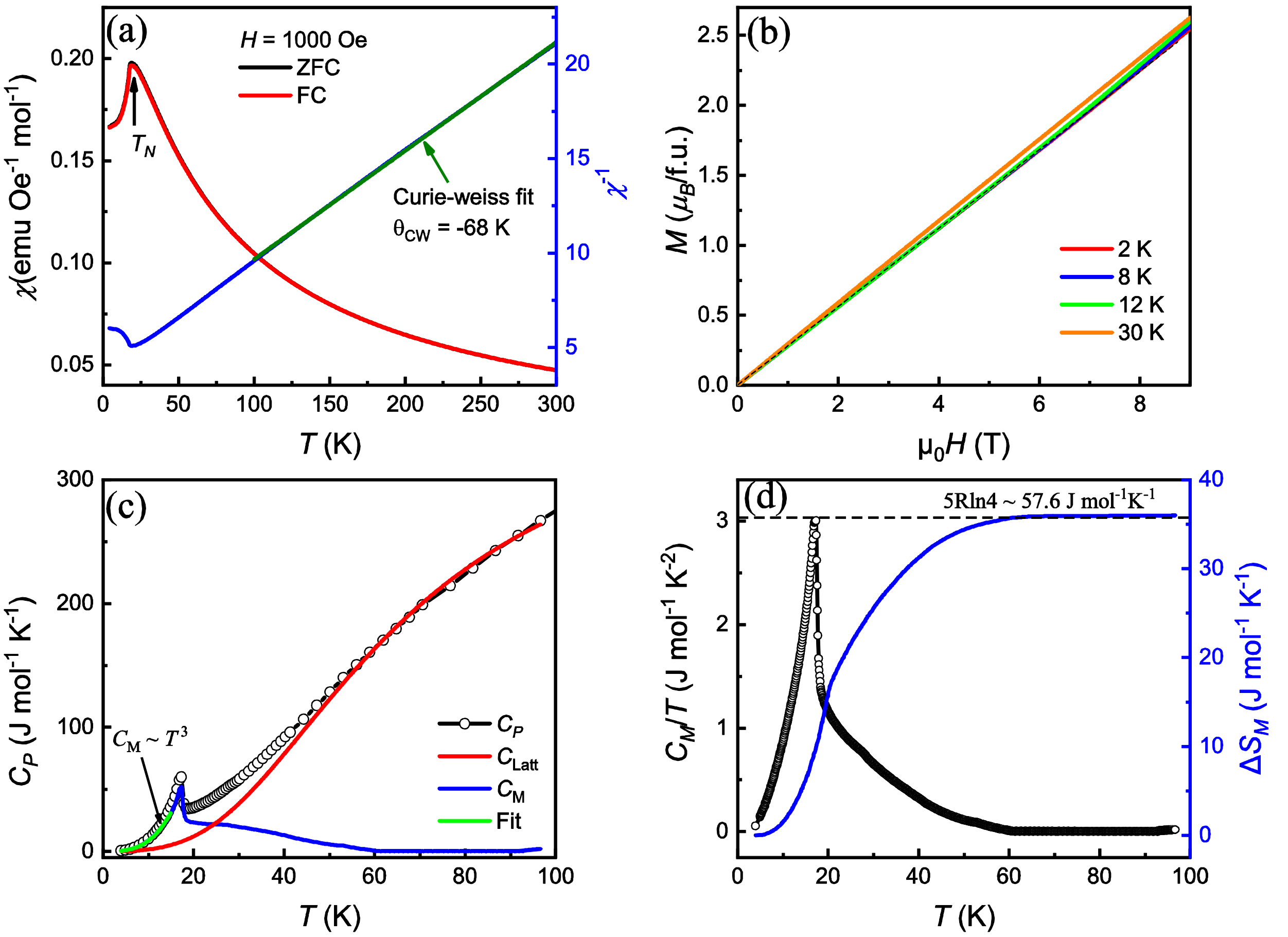

Fig. 2. (a) The $T$ dependence of the magnetic susceptibility of $Co_5(TeO_3)_4Cl_2$ under ZFC and FC modes with measuring field $\mu_0 H$ = 1 kOe. The right axis shows the $T$ dependent inverse susceptibility. (b) Depiction of the isothermal magnetization as a function of $H$ at different temperatures from 2 to 30 K. (c) Heat capacity *vs* temperature curve measured from $T$ = 4 to 100 K. The red line and the blue line represent the lattice contribution ($C_{\mathrm{Latt}}$) and the magnetic contribution ($C_M$), respectively. (d) The $C_M/T$ (left axis) and magnetic entropy ($\Delta S_M$) (right axis) as a function of temperature.

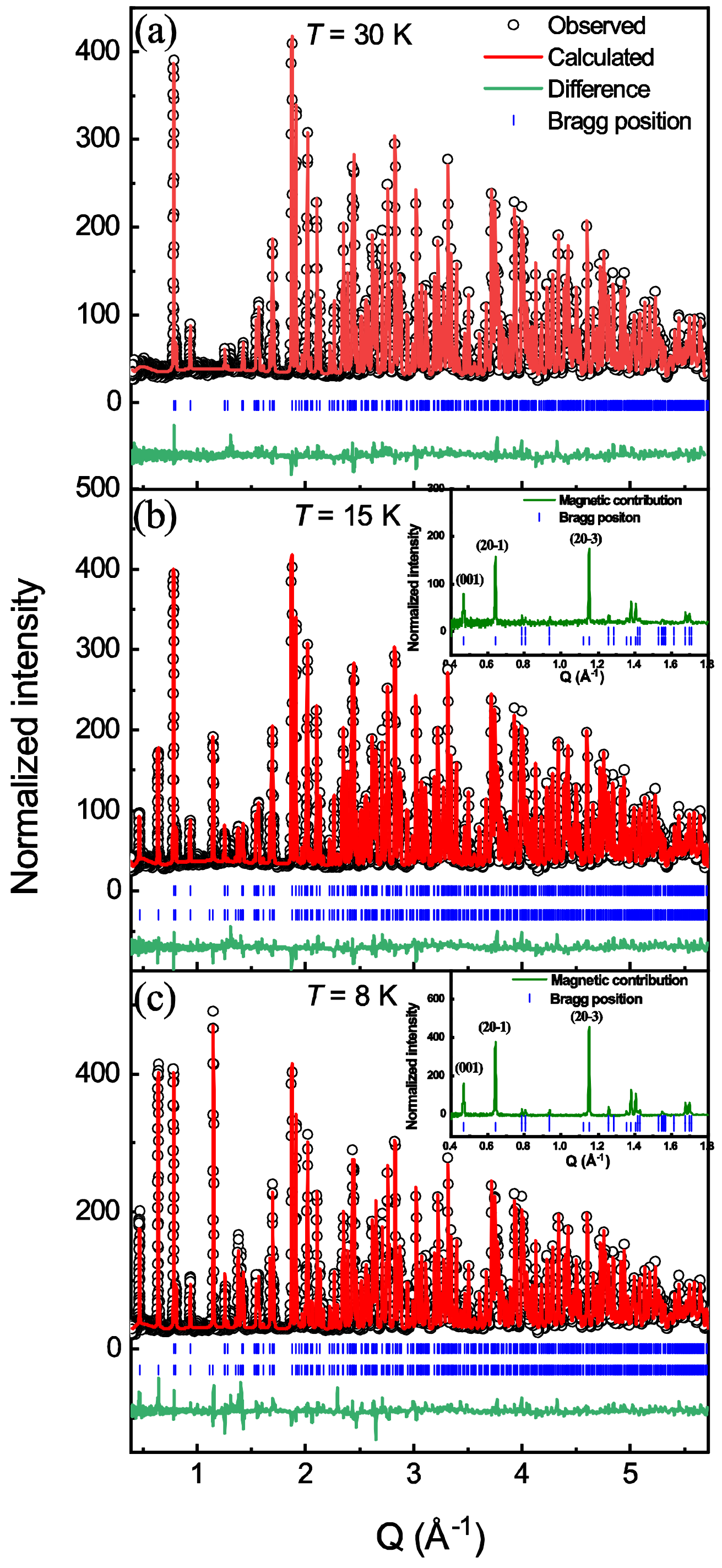

Fig. 3. Rietveld refinement plots of the NPD data at (a) 30 K, (b) 15 K, and (c) 8 K. The ticks below the graphs indicate the calculated Bragg position of the crystal structure (upper row) and magnetic structure (lower row). The magnetic contribution at 15 K and 8 K are shown in the inset.

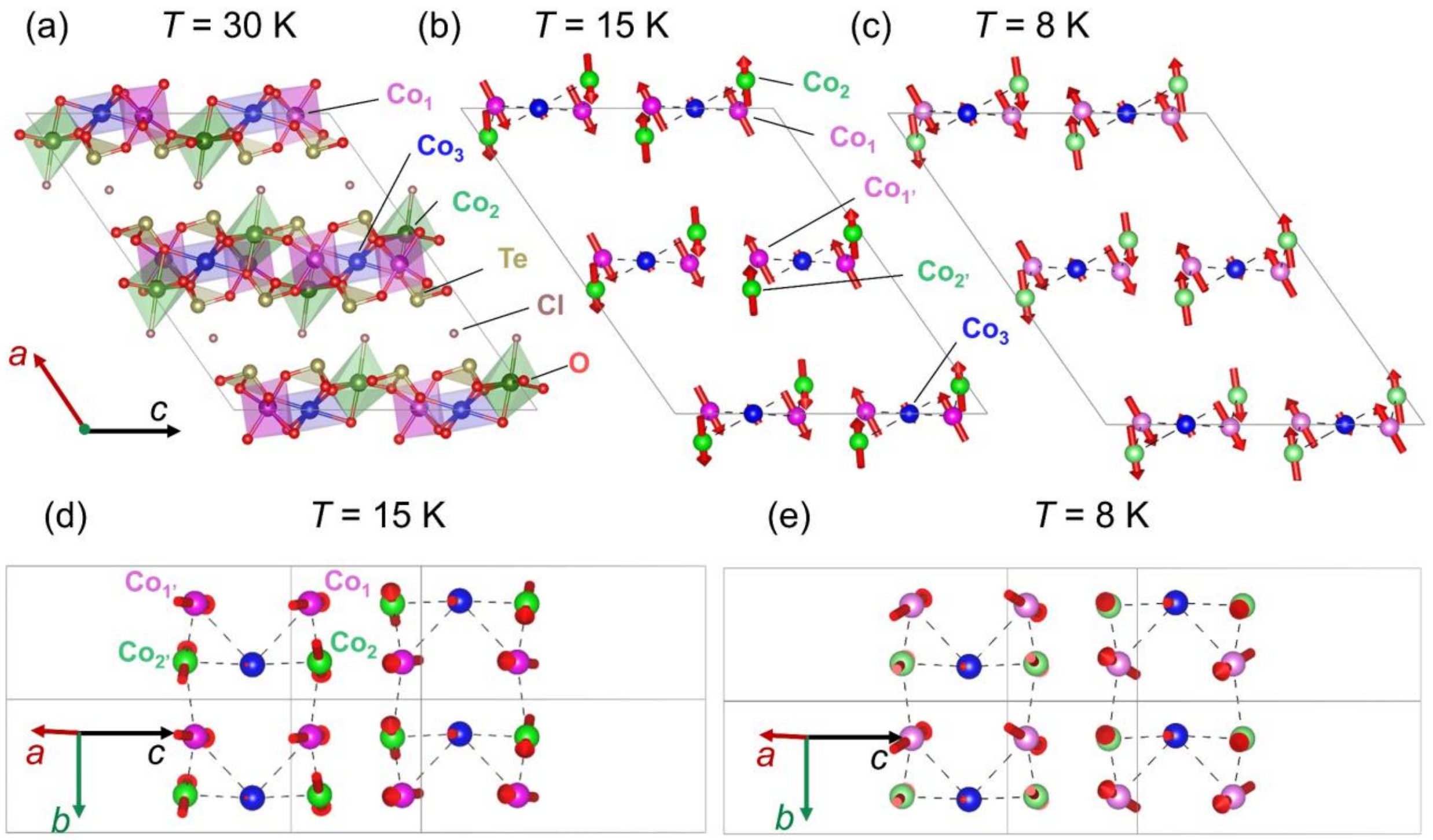

Fig. 4. (a) Illustrations of the crystal structure at 30 K. Schematic view of magnetic structures in the (b) *ac* projection and (d) *bc* projection at 15 K, and (c) *ac* projection and (e) *bc* projection at 8 K.

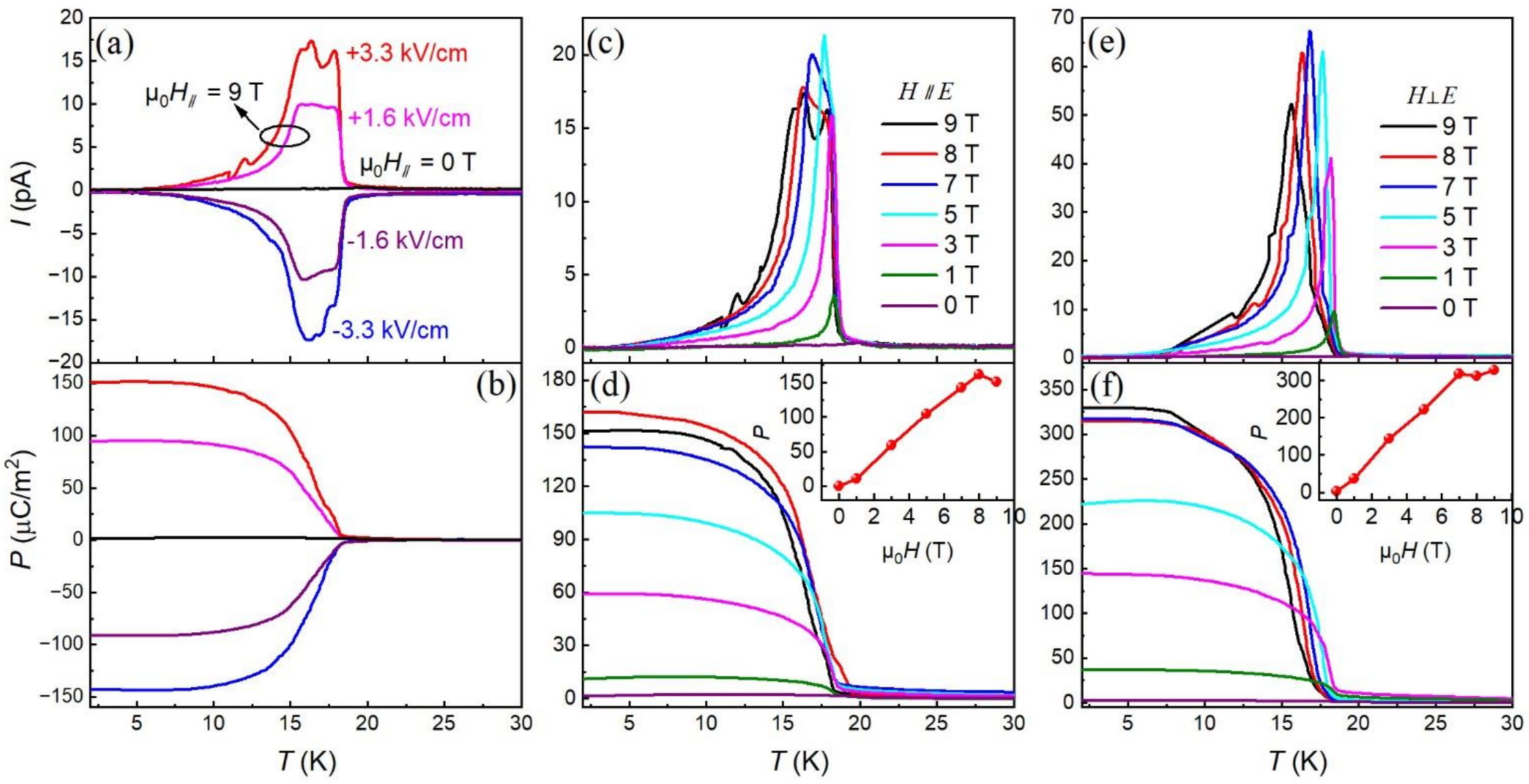

Fig. 5. (a) The pyroelectric current and (b) the electric polarization as a function of temperature collected at $\mu_0 H_{//}$ = 9 T under different poling electric field. The black line shows the measured data at absence of magnetic field (c) The pyroelectric current and (d) the electric polarization as a function of temperature under different magnetic fields in the *H* // *E* configuration. The inset shows the *H* dependent polarization at 2 K. (e) The pyroelectric current and (f) the electric polarization with the magnetic field perpendicular to the electric field. Inset: polarization curves with magnetic field at 2 K.

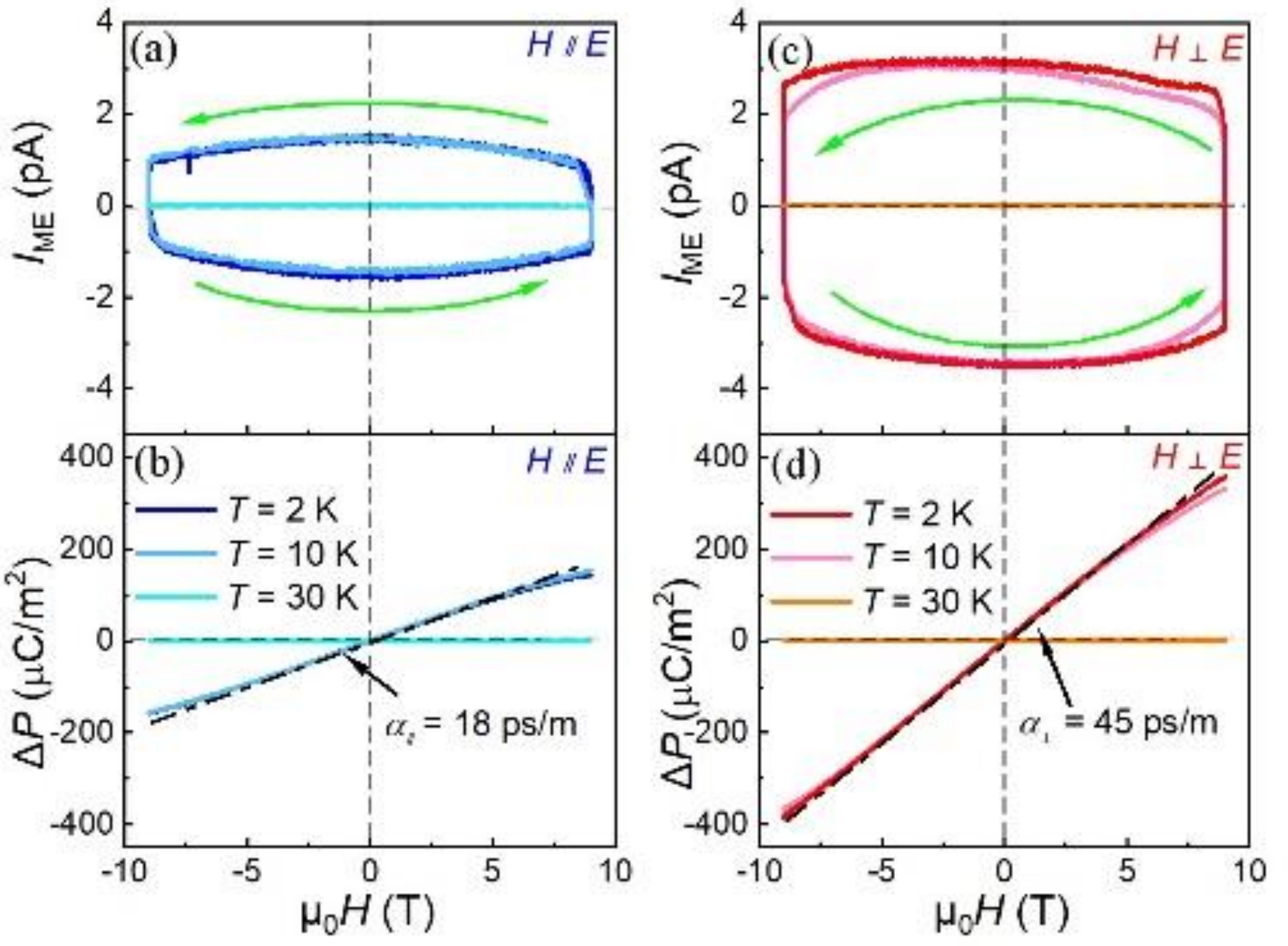

Fig. 6. The ME current $I_{ME}$ as a function of $H$ with (a) $H // E$ and (c) $H \perp E$ measured at $T$ = 2, 10, and 30 K, respectively. The green arrow indicates the direction of the scanning field. The magnetic field dependent polarization $\Delta P$ in the (b) $H // E$ and (d) $H \perp E$ configuration measured at $T$ = 2, 10, and 30 K, respectively. The black dash-dot lines in the (b) and (d) serve as guides to the eye, indicating the linear relationship.

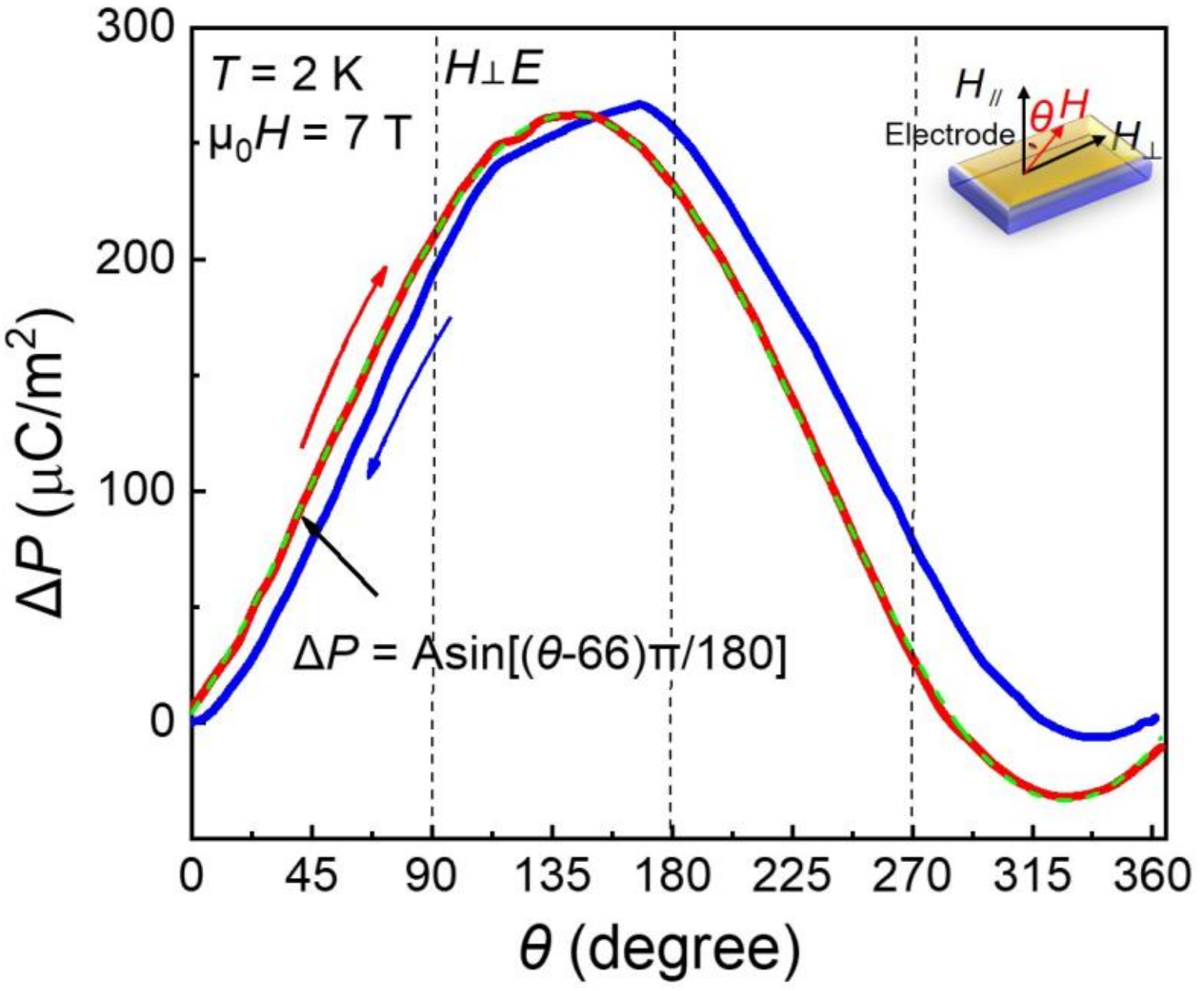

Fig. 7. Angular dependence of polarization under the rotating $\Delta\mu_0 H$ = 7 T at $T$ = 2 K. The arrow indicates the direction of the scanning field (red arrow: 0°~360°, blue arrow: 360°~0°). Schematic sample configuration is shown in the upper right corner.

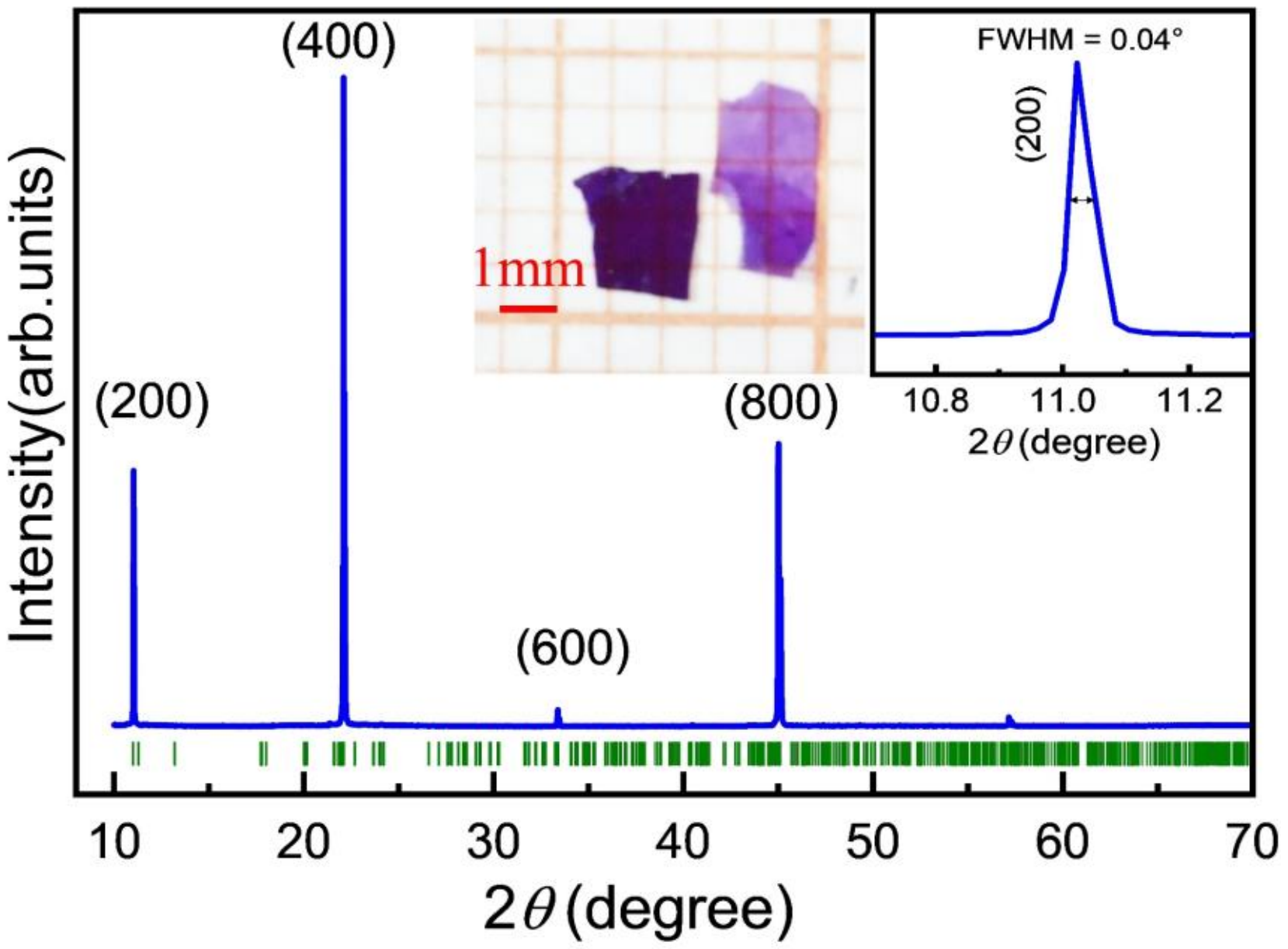

Fig. 8. The XRD pattern of the naturally grown large scale plane of $Co_5(TeO_3)_4Cl_2$ single crystal. Inset shows an optical image of a typical bulk crystal and the full width at half maximum (FWHM) of the (200) diffraction peak.

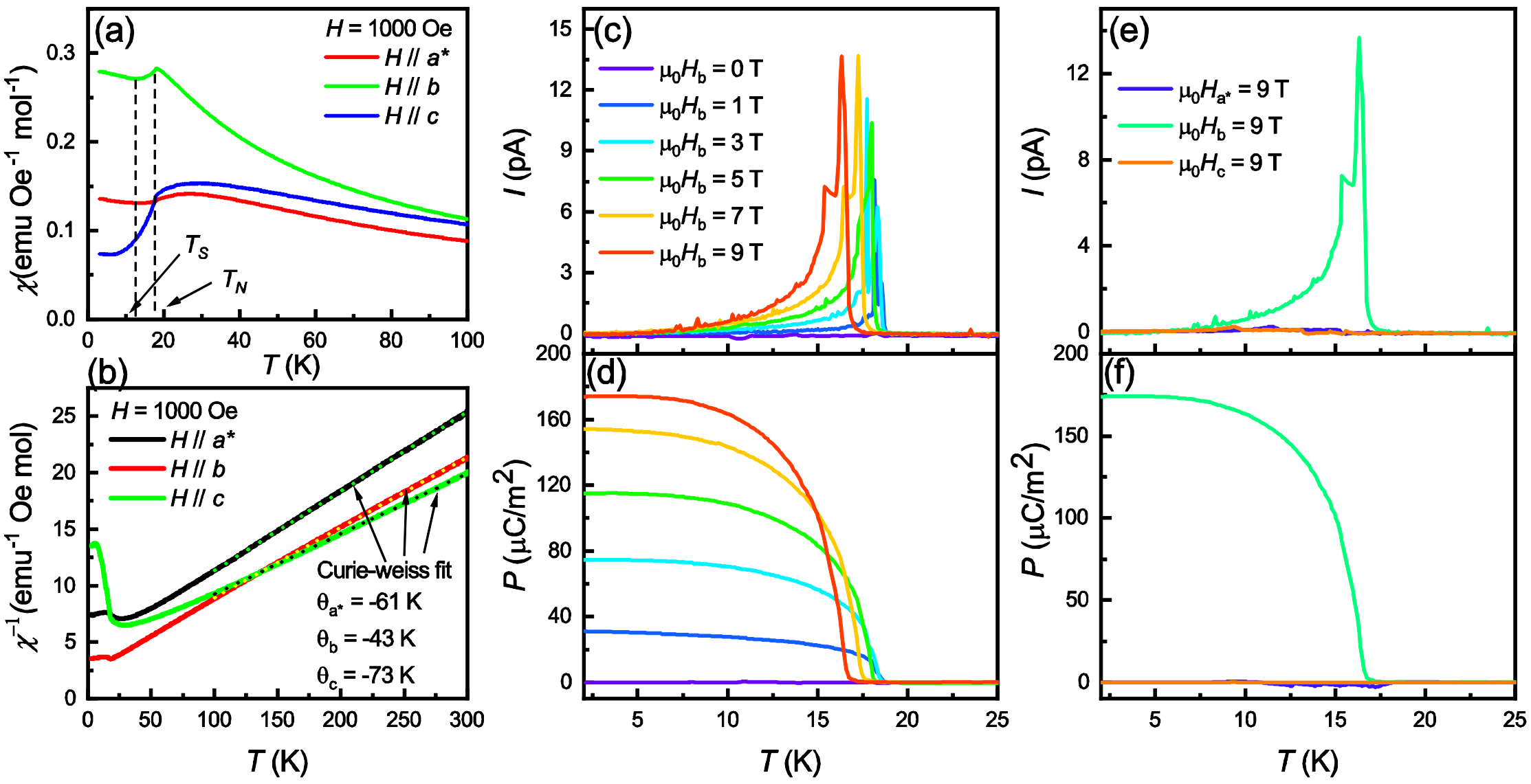

Fig. 9. (a) The $T$ dependence of the magnetic susceptibility of $Co_5(TeO_3)_4Cl_2$ single crystal under ZFC mode with measuring field $\mu_0H$ = 1 kOe along the different crystallographic axes. (b) The $T$ dependent inverse susceptibility and the Curie-Weiss fitting of the data from 100 K to 300 K. (c) The pyroelectric current ($I$) and (d) the electric polarization ($P$) as a function of temperature collected under different magnetic fields $H$ // $b$ axis. (e) The pyroelectric current and (f) the electric polarization as a function of temperature collected for $H$ under different directions with $\mu_0H$ = 9 T.